\documentclass[%
reprint,
 amsmath,amssymb,
 aps,
prb,
floats
]{revtex4-2}

\usepackage{graphicx}
\usepackage{dcolumn}
\usepackage{bm}
\usepackage{xcolor}

\usepackage{braket}
\usepackage{float}
\begin{document}

\title{Extent of frustration in the classical Kitaev-$\Gamma$ model via bond anisotropy}

\author{Ahmed Rayyan$^{1}$, Qiang Luo$^{1}$, Hae-Young Kee$^{1,2}$}
\email{hykee@physics.utoronto.ca}

\affiliation{{\footnotesize{}$^{1}$Department of Physics and Center for Quantum
Materials, University of Toronto, 60 St. George St., Toronto, Ontario,
M5S 1A7, Canada }}
\affiliation{{\footnotesize{}$^{2}$Canadian Institute for Advanced Research, Toronto,
Ontario, M5G 1Z8, Canada}}
\date{\today}

\begin{abstract}
In the pseudospin-$\frac{1}{2}$ honeycomb Mott insulators with strong
spin-orbit coupling, there are two types of bond-dependent exchange
interactions, named Kitaev ($K$) and $\Gamma$, leading to strong frustration.
While the ground state of the Kitaev model is a quantum spin liquid
with fractionalized excitations, the ground state of the $\Gamma$ model
remains controversial. In particular, the phase diagram of the K$\Gamma$
model with ferromagnetic $K$ and antiferromagnetic $\Gamma$ interactions
has been intensively studied because of its relevance to candidate
materials such as $\alpha$-RuCl$_{3}$. Numerical studies also included
the effects of tuning the bond strengths, i.e., $z$-bond strength
different from the other bonds. However, no clear consensus on the
overall phase diagram has been reached yet. Here we study the classical
K$\Gamma$ model with anisotropic bond strengths using Monte Carlo
simulations to understand the phases that emerge out of the competition between
two frustrated limits. We also address how the anisotropic bond strength
affects the phase diagram and strength of quantum fluctuations. We
found various large unit cell phases due to the competing frustrations,
and analyzed their intrinsic degeneracy based on the symmetry of the
Hamiltonian. Using the linear spin wave theory we showed that the
anisotropic bond strength enhances quantum fluctuations in the $\Gamma$-dominant regime where a small reduced moment is observed. The implications
of our findings in relation to the quantum model are also discussed.
\end{abstract}

\maketitle


\section{Introduction}
The Kitaev spin model on the two-dimensional honeycomb lattice serves
as a fascinating example of a quantum spin liquid (QSL) \citep{kitaev2006}.
In particular, the braiding statistics of fractionalized Majorana
excitations in the Kitaev spin liquid (KSL) has generated intense
interest in both condensed matter physics and quantum information
communities due to their application in fault-tolerant quantum computation
\citep{Kitaev_2003}. A key ingredient of the model is a particular
type of bond-dependent interactions resulting in spin frustration,
different from more traditional approaches based on geometrical constraints
or going beyond nearest-neighbor interactions on bipartite lattices
\citep{wannier50,ANDERSON1973153,ANDERSON1196,balents2010spin}. This
intriguing model had been a pure theoretical interest until the Jackeli-Khaliullin
mechanism \citep{jk2009prl,cjk2010prl} outlined how the Kitaev
interactions are generated in the low-energy description of pseudospin
$J_{\text{eff}}=1/2$ moments in spin-orbit coupled Mott insulators.
However, it was shown that in solid-state materials non-Kitaev interactions
are inevitable, and a nearest-neighbor generic model includes another
bond-dependent off-diagonal exchange term named the $\Gamma$ interaction
in addition to a conventional Heisenberg ($J$) term \citep{rau2014prl}.
The generic model was studied using a 24-site exact diagonalization
(ED) which showed a rich phase diagram including various ordered and
disordered phases, but the nature of the disordered phases near the $\Gamma$
region was not identified \citep{rau2014prl}.

A considerable amount of theoretical efforts has been made to pin down
the phase diagram of the extended model and to identify the potential
QSL in Kitaev candidate materials such as $\alpha$-RuCl$_{3}$ \citep{plumb14,hskim15,koitzsch16,sandilands16,zhou16,Banerjee_2016,kim16,winter2016,Rau_2016,Janssen_2017,Winter_2017,Wang_2017,hermanns2018,Takagi2019review,Laurell2020}.
While the KSL and various ordered phases are uncovered, there still
remains regions of the phase space which are not well understood,
with the most peculiar region being that of ferromagnetic (FM) Kitaev
and antiferromagnetic (AFM) $\Gamma$ interactions (K$\Gamma$ model).
Several numerical simulations of the K$\Gamma$ model have reported
quantum disordered phases \citep{yb2018signatures,Gordon_2019,Wang_2019-vmc,Gohlke_2020,Lee2020Magnetic,luo2020spontaneous,Wang_2020_vmc,zhang2021variational},
but the phase diagram is still controversial \citep{Rusnacko2019}.
The classical K$\Gamma$ model in a small phase space near the pure
Kitaev \citep{baskaran2008,Chandra_2010,Rousochatzakis_2018,lampenkelley2021fieldinduced,Chern_2020,sun2021effective}
or pure $\Gamma$ region was also studied \citep{Rousochatzakis_2017,Janssen_2017,Samarakoon_2018,Saha_2019, luo2021gapless,luo2020spontaneous}.
They revealed the macroscopic degeneracy at the pure Kitaev and $\Gamma$
limits \citep{Liu_2021,rao2021machinelearned}, and the large unit cells (LUCs) that cannot be captured by
small clusters used in, for example, ED on the 24-site cluster.

These studies have focused on the isotropic limit, where the exchange
interactions are equivalent on each honeycomb lattice bond. In parallel,
the effects of exchange anisotropy on the spin frustration have also been
explored to find possible QSLs and to understand their connection
to the KSL \citep{catuneanu2018path,Wachtel_2019,Yamada_2020,Wang_2020-vmc}.
They suggest that the strong $z$-bond region hosts large regions
of disordered phases but it is not clear whether they correspond to the
isolated dimer limit \citep{Wachtel_2019,Yamada_2020} or spin liquid
states such as the $\Gamma$ spin liquid ($\Gamma$SL) \citep{luo2020spontaneous,luo2021gapless}
or multi-node gapless QSLs \citep{Wang_2020-vmc}. These numerical
studies may also suffer from finite-size effects and thus an investigation
of the classical K$\Gamma$ model whereby the bond strength is tuned
would offer an insight to the ground states of the anisotropic K$\Gamma$
quantum model. 

In this paper, we tackle this problem by addressing the following
questions. What types of magnetic orderings appear via the competition
between two extreme frustrated limits, i.e., Kitaev and $\Gamma$
limits? How does the exchange anisotropy affect the classical ground
states of the K$\Gamma$ model, and which regions of the anisotropic
phase space may exhibit a quantum-disordered ground state? Using classical
Monte Carlo simulations, we found various LUCs with intriguing fourfold
or eightfold degeneracy except for a few special points with macroscopic
degeneracy. It is likely that LUCs are results of the competition
between the two frustrated Kitaev and $\Gamma$ limits \citep{Chern_2020,Liu_2021}.
Near the $\Gamma$-dominant region, the bond strength anisotropy further
enhances the quantum fluctuations leading to a complete destruction
of the magnetic moment suggesting possible QSLs in this region.

The rest of the paper is organized as follows. In Sec. \ref{sec:Hamiltonian-and-Phase}
we briefly discuss the physics of an isolated $z$-bond before introducing
$x$-and $y$-bond interactions, and then present the phase diagram
of the two-dimensional model obtained via classical Monte Carlo simulations.
In Sec. \ref{sec:Classical--Degenerate} we introduce three symmetry
operations that map each bond Hamiltonian to itself, which reveals
the degeneracy of each ordered phase independent of the bond anisotropy.
In Sec. \ref{sec:SL-and-Related} we focus on the phases near the
$\Gamma$-dominant region, which arise from freezing the Ising degrees
of freedom that form the classical $\Gamma$SL. In Sec. \ref{sec:LSW}
we discuss the effects of quantum fluctuations using linear spin wave
theory (LSWT) \citep{Maestro_2004}. We then summarize our results
and discuss implications of our findings on the quantum model in the
last section.

\section{Dimer Hamiltonian and Classical Phase Diagram\label{sec:Hamiltonian-and-Phase}}
\begin{figure}
\includegraphics[scale=0.22]{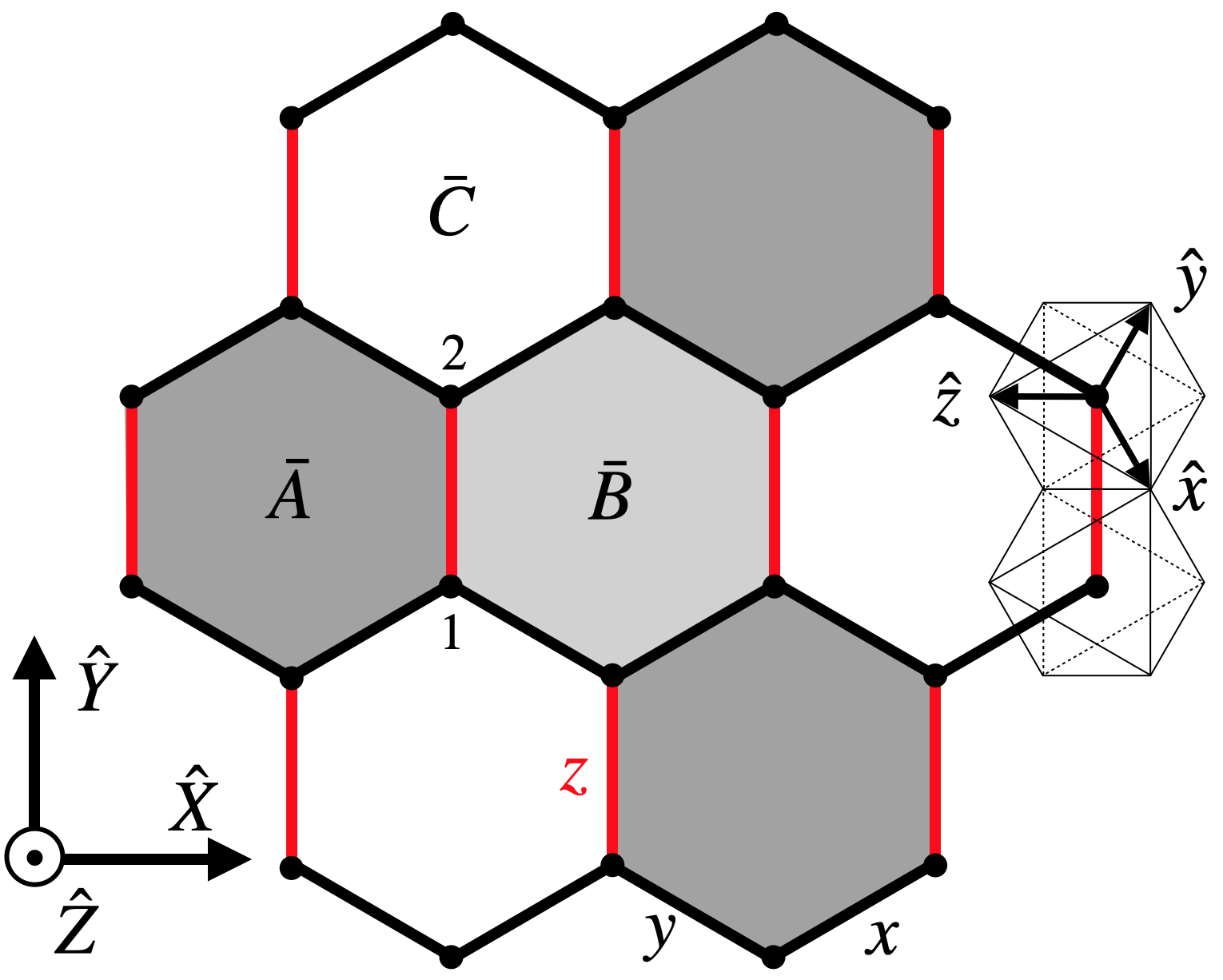}\caption{The honeycomb lattice in the strong $z$-bond limit $\left(g>0\right)$ where the enhanced $z$-bond interaction strength is indicated in red and the $xy$-chain extends horizontally. The three plaquette sublattices $\bar{A},\bar{B},\bar{C}$ are highlighted in dark gray, light gray, and white, respectively. The cubic $xyz$ and crystallographic $XYZ$ bases are also shown, where $\hat{Z}$ is perpendicular to the honeycomb plane and $\hat{X},\hat{Y}$ are perpendicular and parallel to the $z$-bond direction, respectively.}
\label{zbondbasis}
\end{figure}
We study the K$\Gamma$ model on the honeycomb lattice
with bond anisotropy, where the Hamiltonian is given by 
\begin{align}
H & =\sum_{\langle ij\rangle_{\gamma}}K^{\gamma}S_{i}^{\gamma}S_{j}^{\gamma}+\Gamma^{\gamma}\left(S_{i}^{\alpha}S_{j}^{\beta}+S_{i}^{\beta}S_{j}^{\alpha}\right),\label{eq:ham}
\end{align}
where $\gamma\in\left\{ x,y,z\right\} ,\,\alpha,\beta\in\left\{ x,y,z\right\} \backslash\gamma$ label the interaction along a particular bond.
Here the spin direction $\vec{S}_{i}$ is defined in the local octahedral
$xyz$ basis as shown in Fig. \ref{zbondbasis}. The crystallographic
$XYZ$ basis is also shown, where $\hat{Z}=\frac{1}{\sqrt{3}}\left(1,\,1,\,1\right)$
is perpendicular to the honeycomb plane, and $\hat{X}=\frac{1}{\sqrt{6}}\left(1,\,1,-2\right)$,
$\hat{Y}=\frac{1}{\sqrt{2}}\left(-1,\,1,\,0\right)$ are perpendicular
and parallel to the $z$-bonds respectively. These will be used when
we describe the magnetic ordering moment directions in Sec. \ref{sec:Classical--Degenerate}.
$\bar{A},\bar{B},\bar{C}$ denoted by varying shades of gray in Fig.
\ref{zbondbasis} represent the three $\sqrt{3}\times\sqrt{3}$ plaquette
sublattices used in Secs. \ref{sec:Classical--Degenerate} and
\ref{sec:SL-and-Related}. The bond anisotropy is tuned by a parameter
$g\in[0,\,1]$ by introducing the interaction strengths as
\begin{align}
K^{xy}&=-\left(1-g^{2}\right)K,\text{\quad} & K^{z}&=-K,\nonumber\\
\Gamma^{xy}&=+\left(1-g^{2}\right)\Gamma,\ensuremath{\quad} & \Gamma^{z}&=+\Gamma,\label{eq:params}
\end{align}
where $K=\text{cos }\psi$ and $\Gamma=\text{sin }\psi$ and $\psi\in[0,\,0.5\pi]$
is chosen to study the region between the FM Kitaev limit $(\psi=0)$
and the AFM $\Gamma$ limit $(\psi=\pi/2)$. 

The $g=1$ limit corresponds to the case of isolated $z$-bonds, whereas
the three bonds are equivalent at the isotropic limit $g=0$. We study
the $\Gamma/K-g$ phase diagram using classical Monte Carlo to identify
possible magnetic orderings, their origins, and competitions. Before
we present the phase diagram, we analyze the dimer limit $g=1$, which
will be useful to understand several ordered phases appearing when
$g\neq1$. At $g=1$ the K$\Gamma$ Hamiltonian is a sum of isolated
$z$-bonds denoted in red in Fig. \ref{zbondbasis}, and for the $\langle12\rangle_{z}$
bond it is given by
\begin{align}
H_{\langle12\rangle_{z}} & =K^{z}S_{1}^{z}S_{2}^{z}+\Gamma^{z}\left(S_{1}^{x}S_{2}^{y}+S_{1}^{y}S_{2}^{x}\right).\label{eq:bondhamiltonian}
\end{align}
In the classical limit $\vec{S}_{i}$ can be parameterized by $S\left(\text{cos}\,\phi_{i}\;\text{sin}\,\theta_{i},\,\text{sin}\,\phi_{i}\;\text{sin}\,\theta_{i},\text{cos }\theta_{i}\right)$
where $\phi_{i}\in[0,2\pi)$ is the azimuthal angle in the $xy$-plane
and $\theta_{i}\in[0,\pi]$ is the polar angle from the $z$-axis
as in Fig. \ref{zbondbasis}. The bond energy is minimized when $\theta_{0}=\theta_{1}$
and $\phi_{1}+\phi_{2}=-\pi/2$ and the moments can be written as
$\vec{S}_{1}=S\,(a,b,c),\;\vec{S}_{2}=S\,(-b,-a,\,c)$ where $a,b,c\in\mathbb{R}$
satisfy $a^{2}+b^{2}+c^{2}=1$. The bond energy for this configuration
is then
\begin{align}
E_{\langle12\rangle_{z}}/S^{2} & =-\Gamma-\left(K-\Gamma\right)c^{2}.\label{eq:bondenergy}
\end{align}
Note that when $K=\Gamma$ $\left(\psi=\pi/4\right)$ each of the
$N/2$ isolated $z$-bonds retain an $O(3)$ symmetry, where $N$
is the number of sites. Away from this point the $O(3)$ symmetry
is lifted and one of two states may stabilize, while the macroscopic
degeneracy associated with each $z$-bond remains. When $K>\Gamma$
the energy is minimized by setting $c=1$ and the moments form FM
dimers pinned along the $\hat{z}$ direction with a twofold Ising degeneracy.
On the other hand, when $\Gamma>K$ the bond energy is minimized by
placing the moments in the $xy$-plane. In this case the moments retain
a continuous $O(2)$ degeneracy with the restriction that $\phi_{1}+\phi_{2}=-\pi/2$.
The transition between the two phases is thus a first-order spin-flop
transition. Introducing interactions along the $x-$ and $y$-bonds
when $g\neq1$ may lift the macroscopic degeneracy and one may wonder
what possible orderings arise from the peculiar point, and how far
they can be extended, i.e., if they can survive all the way to the
two-dimensional isotropic limit. To answer these questions, we solve
the classical model numerically using simulated annealing Monte Carlo
(SAMC) \citep{metropolis53,kirkpatrick1983optimization,kirkpatrick1984optimization}
on clusters of up to $N=720$ sites with $N\times10^{5}$ MC steps,
see Appendix \ref{sec:Simulation-Details} for simulation details.
\begin{figure}
\includegraphics[scale=0.53]{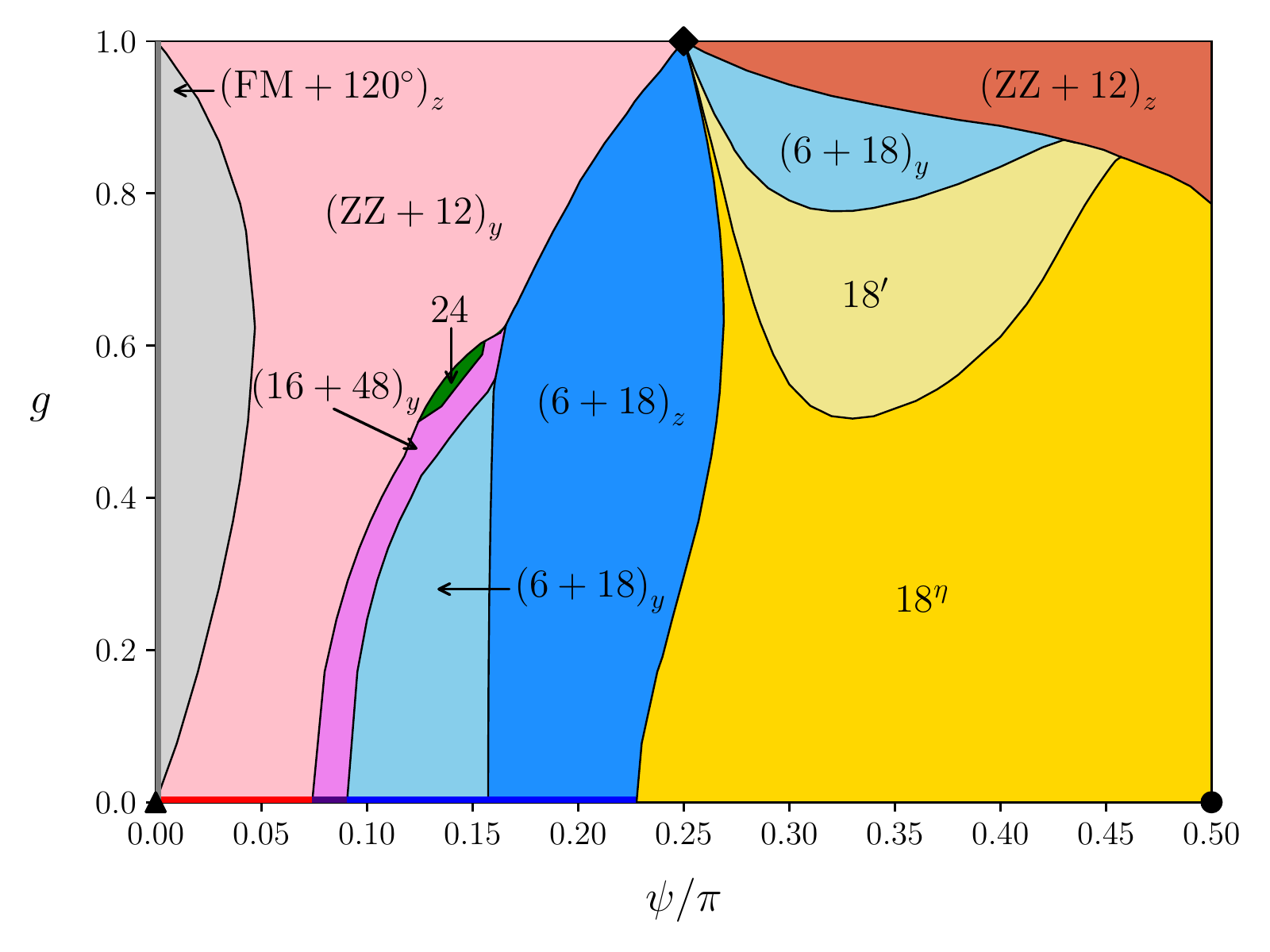}\caption{The classical phase diagram of the anisotropic K$\Gamma$ model. All phases are separated by first-order transitions except for the $18^{\eta}$ and $18'$ phase boundary, which is of second-order. We highlight three points with macroscopic
degeneracy, the $K=\Gamma$ dimer point and the pure isotropic $K$
and $\Gamma$ points, with a black diamond, triangle, and circle,
respectively. In the isotropic limit the three orientations of {\color{black}$\text{ZZ}+12$, 
$6+18$, and $16+48$ phases are degenerate, which we indicate by the
dark red, dark blue, and dark purple lines along $g=0$, respectively.} This degeneracy is lifted
for $g\protect\neq0$ due to the broken $C_{3}$ symmetry. At $\psi=0$
we indicate the Kitaev FM dimer phase in dark gray, which immediately
forms the $\text{FM}+\text{120}^{\circ}$ order in the presence of
$\Gamma>0$. See the main text for the ordering patterns of the phases
and their SSF peaks in Appendix \ref{sec:SSFclassical}.}
\label{pd}
\end{figure}

We present the phase diagram in Fig. \ref{pd}. The phase boundaries
are obtained by comparing the energies of the classical states, and
the nature of the phase transitions across the boundaries is determined
by the first singular derivative of the ground state energy per site
$E_{0}/N$. The presence of magnetic order can be identified by sharp
features in the static structure factor (SSF) $s_{\mathbf{k}}=\sum_{ij}\vec{S}_{i}\cdot\vec{S}_{j}\,e^{i\mathbf{k}\cdot\left(\mathbf{r}_{i}-\mathbf{r}_{j}\right)}$
where $i,j$ range over all sites of the cluster and $\mathbf{r}_{i}$
is the vector pointing to site $i$. 

The phase diagram shows several spin orderings and in Appendix \ref{sec:SSFclassical}
we show the SSF of each order. Following Ref. \citep{Chern_2020}
we use the notation of $n_{1}+n_{2}$, where ``$+$'' represents
the degeneracy of two orderings and $n_{i}$ denotes the
number of sites in the phase's magnetic unit cell. Three exceptions are the FM,
zigzag (ZZ), and $120^{\circ}$ orders which have two, four, and six sites
in the magnetic unit cell respectively. In Fig. \ref{pd} we find
the presence of these three phases as well as the{ \color{black} six-
and $16$-site orders: the former} appears in Refs. \citep{rau2014trigonal,Chern_2020,lampenkelley2021fieldinduced}
and is composed of alternating zigzag and stripy chains, see Fig.
\ref{618z}(a) in the next section. Interestingly, some of these phases
are degenerate with an order containing a larger unit cell with {\color{black} 12, 18 and 48 sites}. There also exists the $\ensuremath{18^{\eta}}$,
$18'$, and 24-site orders which do not have a smaller ordering counterpart.
We note briefly that the LUC orders generally contain
dominant SSF peaks at multiple wavevectors within the $1^{\text{st}}$
(crystal) Brillouin zone. In particular the $18^{\eta}$ and $18'$
orders have finite spectral weight at the three $\mathbf{Q}=\frac{2}{3}M,\,\frac{4}{3}M$
points of the crystal Brillouin zone, see Appendix \ref{sec:SSFclassical}.

We distinguish the phases that share the same ordering pattern but
with a different moment orientation using a subscript of $i=x,y,z$.
For example, the $\text{ZZ}_{z}$ and $\text{ZZ}_{y}$ both have four
sites in the magnetic unit cell, but the zigzag chains repeat along
the $z$-bond direction in the former and the $y$-bond direction
in the latter. There is also a $\text{ZZ}_{x}$ orientation that
is degenerate with $\text{ZZ}_{y}$ and can be obtained by a $C_{2}$
rotation about the $z$-bond direction $\hat{Y}$, but it is omitted
in Fig. \ref{pd} for simplicity. 

The pure classical Kitaev model exhibits an extensive ground-state
degeneracy, and when $g>0$ the moments form disconnected FM dimers
along the $z$-bond, which point in the $\pm\hat{z}$ direction \citep{baskaran2008,Chandra_2010}
. This is denoted by the solid gray line in Fig. \ref{pd}. When $\Gamma$
is turned on the magnetic order is stabilized and a FM is formed with
moments pinned near the $\hat{z}$ axis. The $120^{\circ}$ order,
which is degenerate with the FM, can be obtained by a symmetry operation
to be discussed in the next section. In a small region between the
$\left(\text{ZZ}+12\right)_{y}$ and $6+18$ regions we find {\color{black} $16+48$ and $24$-site
orders} which may arise from further moment frustration. Below we will
focus our attention on the $\text{ZZ}+12$, $6+18$, $18^{\eta}$,
and $18'$ phases, which occupy the majority of the phase space extending
from the $K=\Gamma$ dimer point at $\left(\psi,g\right)=(0.25\pi,\,1)$
to the isotropic limit. The $\text{ZZ}+12$ and $6+18$ orders are
stabilized in the Kitaev-dominant region and are sensitive to anisotropy
since a particular orientation is selected based on the values of $g$
and $\Gamma/K$. On the other hand, the $18^{\eta}$ and $18'$ phases
remain eightfold-degenerate throughout their respective phase regions.
In the next two sections we present the patterns of each magnetic
order and the symmetries related to their degeneracy, and in Sec.
\ref{sec:LSW} we will discuss the quantum effects on the magnetically
ordered states using LSWT.

\section{Classical K$\Gamma$ Degenerate Manifolds\label{sec:Classical--Degenerate}}
The degeneracy of the classical orders exhibited in Fig. \ref{pd}
originates from a symmetry of the Hamiltonian Eq. \eqref{eq:ham}.
This can be seen by considering the dual honeycomb lattice, i.e., a
triangular network with sites at the center of each hexagon labeled
by the three plaquette sublattices $\bar{A},\,\bar{B},\,\bar{C}$
in Fig. \ref{zbondbasis}. For a sublattice $\bar{\sigma}\in\left\{ \bar{A},\,\bar{B},\,\bar{C}\right\} $,
we define the operation
\begin{equation}
\mathcal{R}_{\bar{\sigma}}=\prod_{p\in\bar{\sigma}}\prod_{i\in\partial p}C_{2}^{\text{out(\ensuremath{i})}},\label{eq:Rtransform}
\end{equation}
where $C_{2}^{\alpha}$ is a $\pi$ rotation about the cubic $\alpha$-axis
and $\text{out(\ensuremath{i})}=x,y,z$ refers to the bond which extends
outwards from the plaquette $p\in\bar{\sigma}$ at site $i$. The
notation $\partial p$ refers to the boundary of the plaquette $p$,
which consists of six bonds. For example, the $\langle12\rangle_{z}$
bond Hamiltonian transforms under $\mathcal{R}_{\bar{A}}$ as
\begin{align}
H_{\langle12\rangle_{z}} & =K^{z}S_{1}^{z}S_{2}^{z}+\Gamma^{z}\left(S_{1}^{x}S_{2}^{y}+S_{1}^{y}S_{2}^{x}\right)\nonumber \\
 & \xrightarrow{\mathcal{R}_{\bar{A}}}K^{z}\left(-S_{1}^{z}\right)\left(-S_{2}^{z}\right)+\Gamma^{z}\left[\left(-S_{1}^{x}\right)\left(-S_{2}^{y}\right)+S_{1}^{y}S_{2}^{x}\right]\nonumber \\
 & =H_{\langle12\rangle_{z}}.\label{eq:bondRa}
\end{align}
If we apply this operation to the remaining sites of the honeycomb
lattice, the total Hamiltonian Eq. \eqref{eq:ham} maps to
itself. This is the case for the $\mathcal{R}_{\bar{B}},\mathcal{R}_{\bar{C}}$
operators defined for the plaquette sublattices $\bar{B},\bar{C}$,
respectively. Thus the Hamiltonian is intact under the three $\mathcal{R}_{\bar{\sigma}}$
symmetry operations of Eq. \eqref{eq:Rtransform}. Since $\left(\mathcal{R}_{\bar{\sigma}}\right)^{2}=\mathcal{R}_{\bar{A}}\mathcal{R}_{\bar{B}}\mathcal{R}_{\bar{C}}=1$,
the set $\left\{ 1,\,\mathcal{R}_{\bar{A}},\,\mathcal{R}_{\bar{B}},\,\mathcal{R}_{\bar{C}}\right\} $
is isomorphic to the Klein four-group $\mathbb{Z}_{2}\times\mathbb{Z}_{2}$.
This group leads to the degeneracy in all the phases shown in Fig.
\ref{pd}.

We note that the $\mathcal{R}_{\bar{\sigma}}$ transformations were
first introduced in the context of the pure isotropic $\Gamma$ model
\citep{Rousochatzakis_2017}. In this work we show that these operations
continue to be symmetries of the K$\Gamma$ model with finite $g$.
\begin{figure}
\includegraphics[scale=0.33]{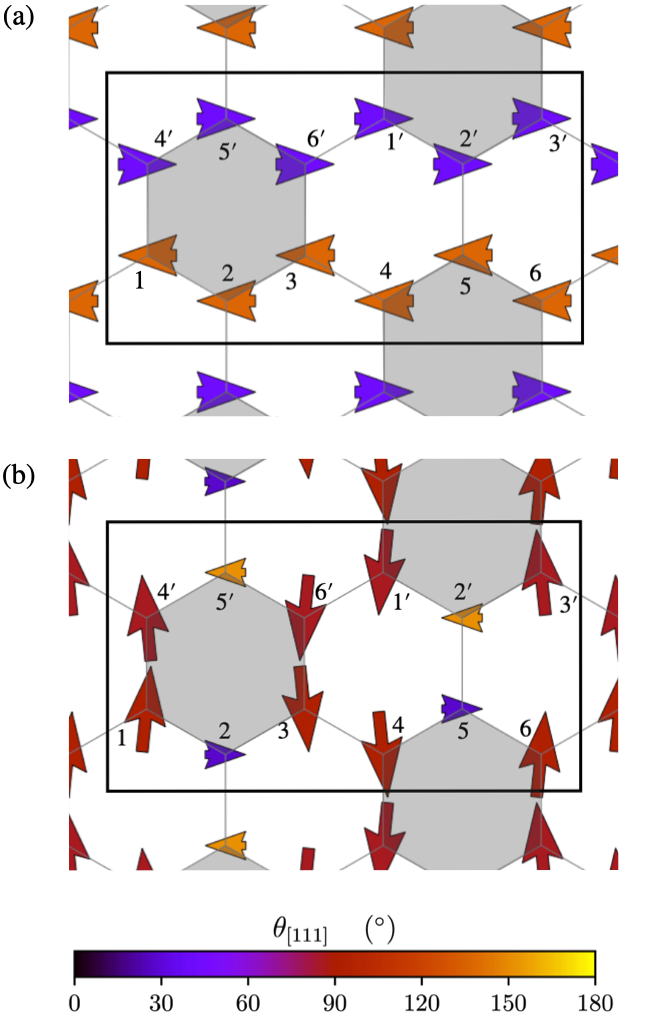}
\caption{The (a) $\text{ZZ}_{z}$ and (b) $\text{12}_{z}$ configurations at
$\left(\psi,g\right)=\left(0.4\pi,0.94\right)$. The plaquette sublattice
$\bar{A}$ is shown in gray. The color of each moment in Figs. \ref{zz12rt}-\ref{18etavzy} denotes the angle made with respect to the $\left[111\right]$ direction.}
\label{zz12rt}
\end{figure}

\begin{figure}
\includegraphics[scale=0.22]{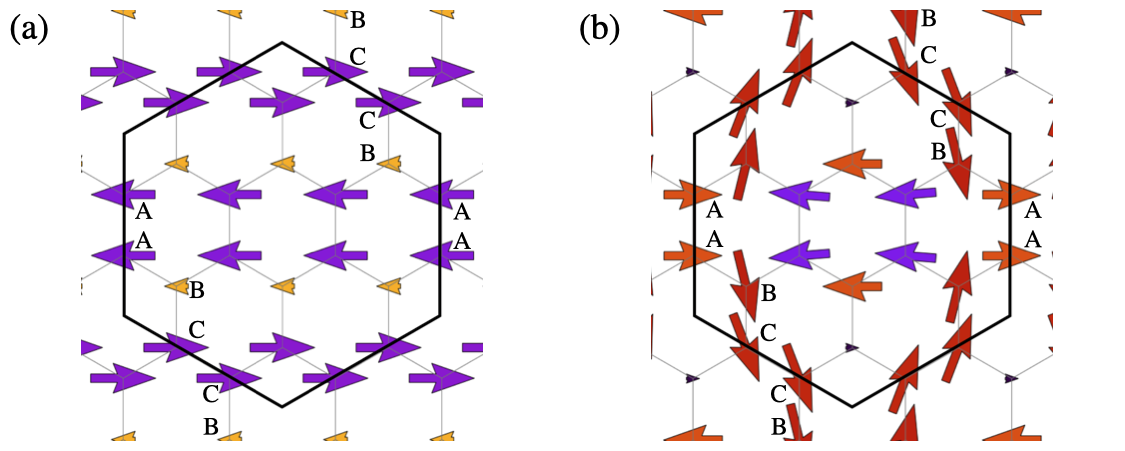}

\caption{The (a) $6_{z}$ and (b) $18_{z}$ configurations at $\left(\psi,\,g\right)=(0.25\pi,\,0.84)$,
where the counterrotating spiral pattern ABCCBA pattern is shown.
Whereas the $6_{z}$ orientation contains a chain of FM aligned moments
along the $x$- and $y$-bonds, in the $6_{y}$ orientation the chain
runs along the $x$- and $z$- bonds.}

\label{618z}
\end{figure}

Let us now explore the action of $\mathcal{R}_{\bar{\sigma}}$ on
the phases shown in Fig. \ref{pd} using the $\left(\text{ZZ}+12\right)_{z}$
orientation as an example. The $\text{ZZ}_{z}$ orientation shown
in Fig. \ref{zz12rt}(a) is given by FM chains of moments $\vec{S}{}_{i}=\hspace{0.27cm}S(-a,-a,+c),\:\vec{S}{}_{i'}=S(+a,+a,-c)$
separated by $z$-bonds, where $|S_{i,i'}^{x}|=|S_{i,i'}^{y}|=S\,a,$
$|S_{i,i'}^{z}|=S\,c$ and $2a^{2}+c^{2}=1$. These moments lie in
the crystallographic $XZ$ plane and $a\gg c$ due to the proximity
to the $\Gamma$ dimer limit, which favours moments lying in the $xy$-plane.
Now we apply the symmetry operation $\mathcal{R}_{\bar{A}}$ on this
configuration where $\bar{A}$ is the plaquette sublattice shown in
dark gray in Fig. \ref{zz12rt}. Explicitly we perform $C_{2}^{x}$
rotation on sites $1,6,1',6'$, $C_{2}^{y}$ on sites $3,4,3',4'$,
and $C_{2}^{z}$ on sites $2,5,2',5'$ which results in the $12$-site
order shown in Fig. \ref{zz12rt}(b) given by
\begin{align}
\vec{S}{}_{1}=S & (-a,+a,-c), & \vec{S}{}_{4'}=S & (-a,+a,+c),\nonumber \\
\vec{S}{}_{2}=S & (+a,+a,+c), & \vec{S}{}_{5'}=S & (-a,-a,-c),\nonumber \\
\vec{S}{}_{3}=S & (+a,-a,-c), & \vec{S}{}_{6'}=S & (+a,-a,+c),\nonumber \\
\vec{S}{}_{4}=S & (+a,-a,-c), & \vec{S}{}_{1'}=S & (+a,-a,+c),\nonumber \\
\vec{S}{}_{5}=S & (+a,+a,+c), & \vec{S}{}_{2'}=S & (-a,-a,-c),\nonumber \\
\vec{S}{}_{6}=S & (-a,+a,-c), & \vec{S}{}_{3'}=S & (-a,+a,+c).\label{eq:12}
\end{align}
We call this particular orientation $12_{z}$ due to the two-site
periodicity along the $z$-bond as shown in Fig. \ref{zz12rt}(b).
This shows that the two orders are degenerate. The two other operations
$\mathcal{R}_{\bar{B},\bar{C}}$ applied on $\text{ZZ}_{z}$ give
the $12_{z}$ order up to translations of the magnetic unit cell,
so that the total degeneracy due to $\mathcal{R}_{\bar{\sigma}}$
and time reversal $\mathcal{T}:\vec{S}_{i}\rightarrow-\vec{S}_{i}$
is four. Similarly, the $\left(\text{ZZ}+12\right)_{y}$ orientation
is four-fold degenerate but cannot be\emph{ }mapped to the $\left(\text{ZZ}+12\right)_{z}$
orientations when $g\neq0$ and a first order transition separates
the two. This analysis applies to the three orientations
of the $\text{FM}+120^{\circ}$,  {\color{black} $6+18$, and $16+48$ orders} as well: the $\left(6+18\right)_{z}$
orientation is shown in Fig. \ref{618z}.

From Eq. \eqref{eq:12} we note that the moments in the $12_{z}$
configuration alternate in a six-site pattern ABCCBA along the $xy$-chain.
This is also the case for the $6+18$ phases as indicated in Fig.
\ref{618z}. This pattern is referred to as a counterrotating spiral
as the moments alternate as ABC along one site sublattice and ACB
along the other, forming two FM dimers serving as inversion centers.
This pattern has appeared in previous models of hyperhoneycomb materials
\citep{Lee_2015,Kimchi_2016,Ducatman_2018,Stavropoulos_2018} suggesting
a relation between the $\mathrm{12}$ and $6+18$ phases to the so-called
$K$ states of Ref. \citep{Ducatman_2018}.
\section{Freezing the $\Gamma$ Spin Liquid: $18^{\eta}$ and $18'$ Phases
\label{sec:SL-and-Related}}
\begin{figure}
\includegraphics[scale=0.22]{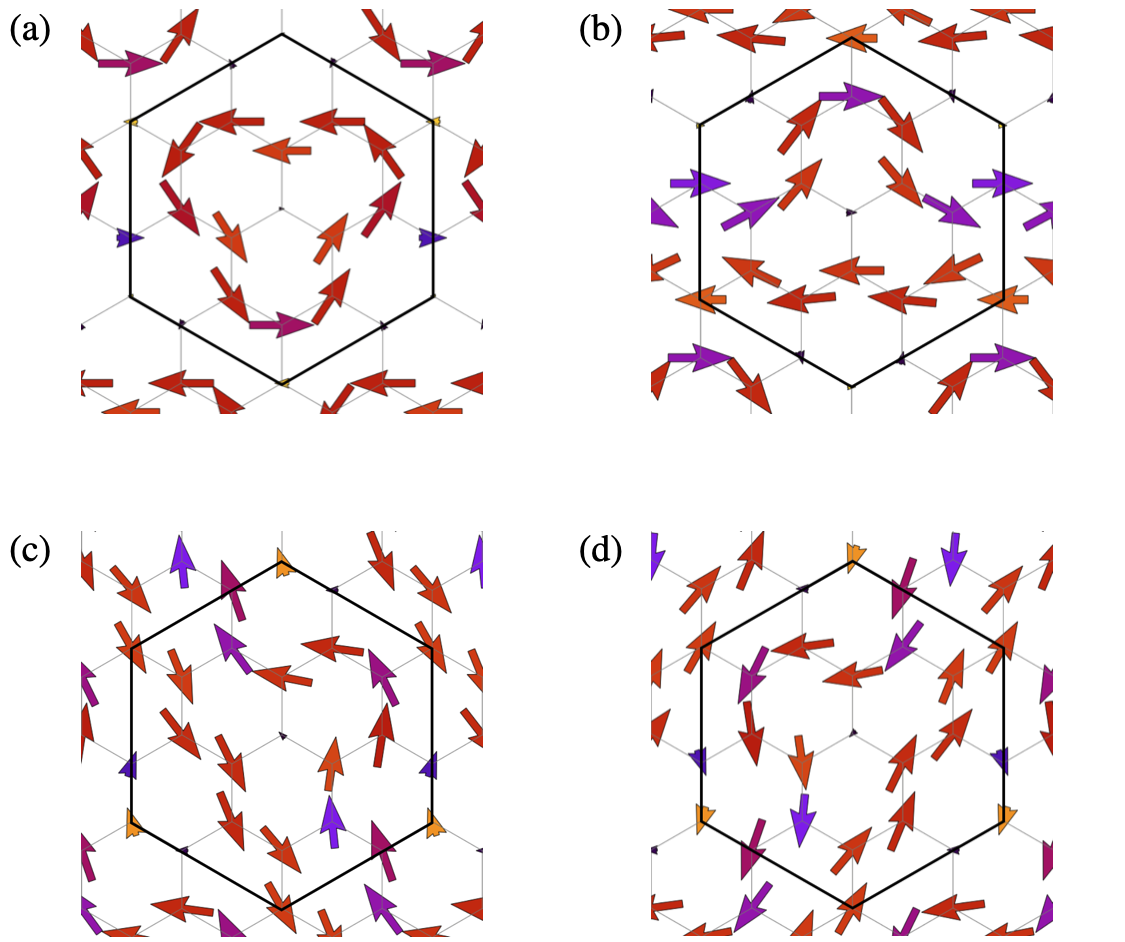}
\caption{The (a) $18_{V}^{\eta}$, (b) $18_{z}^{\eta}$, (c) $18_{x}^{\eta}$,
and (d) $18_{y}^{\eta}$ configurations at $\left(\psi,\,g\right)=\left(0.4\pi,\,0.5\right)$.}
\label{18etavzy}
\end{figure}
In this section we focus on the $18^{\eta}$ and $18'$ phases,
where there are four degenerate spin patterns (excluding time-reversal
partners) with the same size of magnetic unit cell. One vortex pattern
is denoted by $18_{V}^{\eta}$ (called $18$-$C_{3}$ in Ref. \citep{Chern_2020})
and three patterns as $18_{i}^{\eta}$ for $i\in\left\{ x,y,z\right\} $,
see Fig. \ref{18etavzy}. Crucially, the four orientations are connected
by the three $\mathcal{R}_{\bar{\sigma}}$ operations.

\begin{figure}
\includegraphics[scale=0.19]{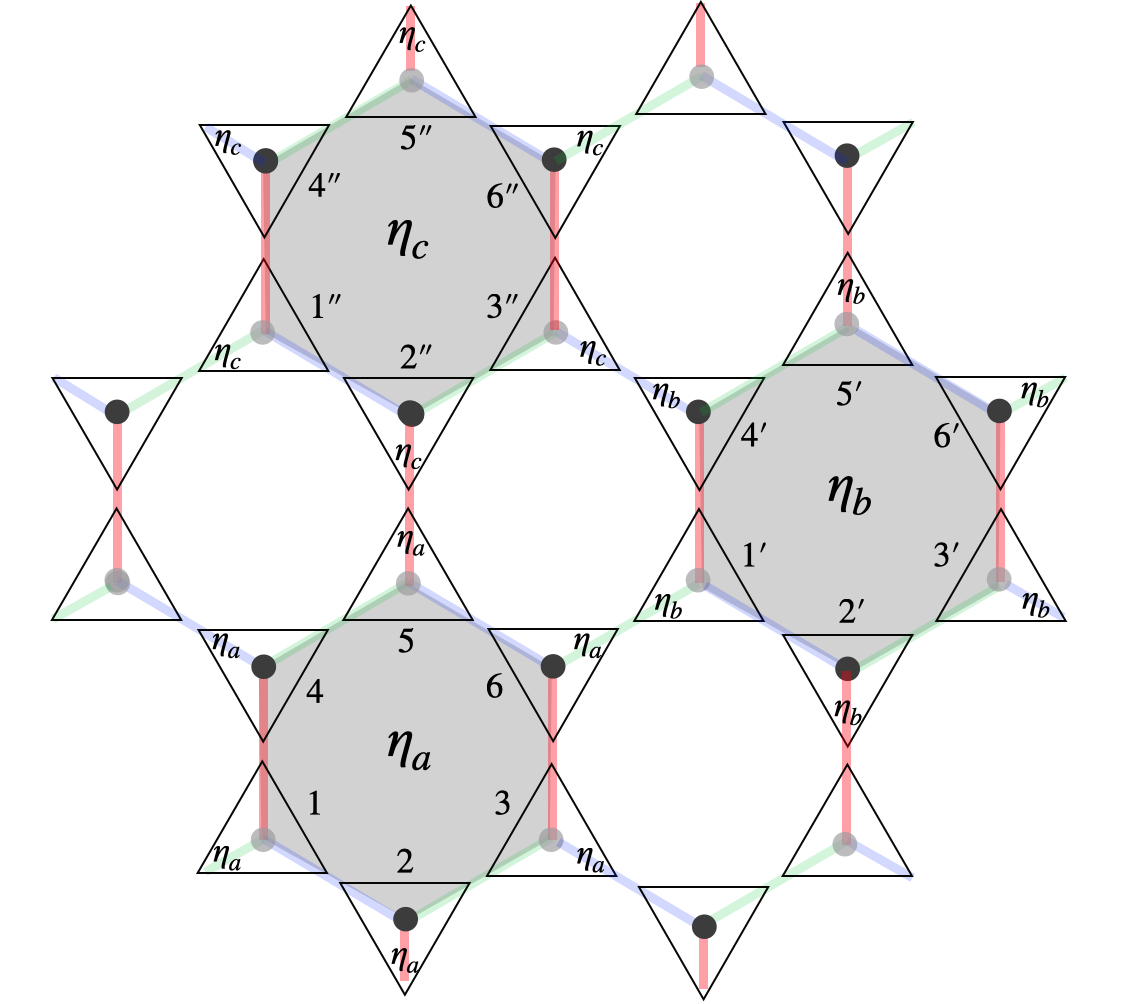}\caption{$\eta$-representation of the moments $\vec{S}_{i}$ in the pure $\Gamma$
limit, where we only display the $\eta_{p}$ that reside on the plaquette
sublattice $\bar{A}$ (shown in gray) for clarity.}

\label{gsleta}
\end{figure}

To discuss the appearance of the $18^{\eta}$ and $18'$ phases in
the $\Gamma$-dominant limit, and the difference between the two,
we first review the physics of the isotropic $\Gamma$ model. There
it was found that the classical ground state is the $\Gamma$ spin
liquid ($\Gamma$SL), which contains an extensive degeneracy due to
free Ising degrees of freedom $\eta_{p}=\pm1$ that reside on each
plaquette $p$ in addition to a continuous $O(3)$ degeneracy \citep{Rousochatzakis_2017}.
This can be seen by separating the sign and magnitude of the spin
components as
\begin{equation}
\vec{S}_{i}=\begin{cases}
+S\left(\eta_{i}^{x}\,a_{i},\,\eta_{i}^{y}\,b_{i},\,\eta_{i}^{z}\,c_{i}\right) & i\ensuremath{\in}\text{A sublattice,}\\
-S\left(\eta_{i}^{x}\,a_{i},\,\eta_{i}^{y}\,b_{i},\,\eta_{i}^{z}\,c_{i}\right) & i\in\text{B sublattice,}
\end{cases}\label{eq:sublattice}
\end{equation}
where A and B are the honeycomb site sublattices and $(a_{i},\,b_{i},\,c_{i})=(|S_{i}^{x}|,\,|S_{i}^{y}|,\,|S_{i}^{z}|)/S$
satisfies $a_{i}^{2}+b_{i}^{2}+c_{i}^{2}=1$ and $\eta_{i}^{\alpha}=\pm1,$
$\alpha\in\left\{ x,\,y,\,z\right\} $. We introduce a visual guide
where each site $i$ is represented by a triangle with each corner
corresponding to one of the three $\eta_{i}^{\alpha}$
\begin{equation}
\vec{\eta}_{i}=\left(\eta_{i}^{x},\,\eta_{i}^{y},\,\eta_{i}^{z}\right)=\begin{cases}
\includegraphics[scale=0.0525]{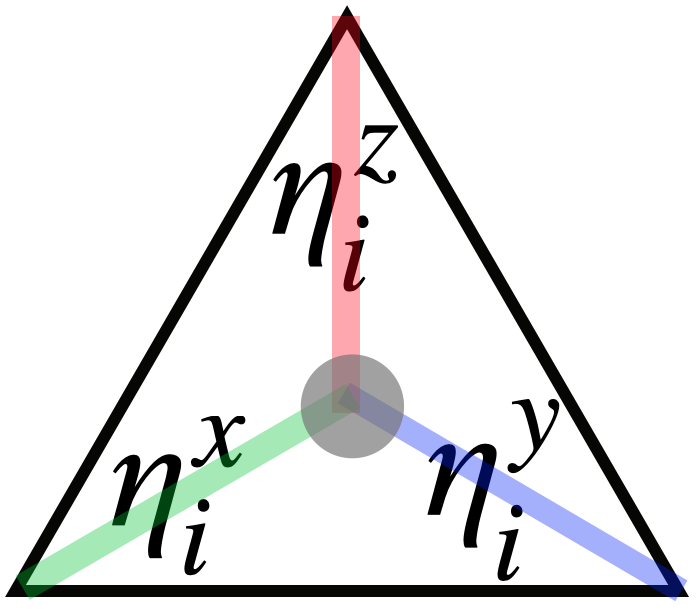} & i\ensuremath{\in}\text{A sublattice,}\\
\includegraphics[scale=0.0525]{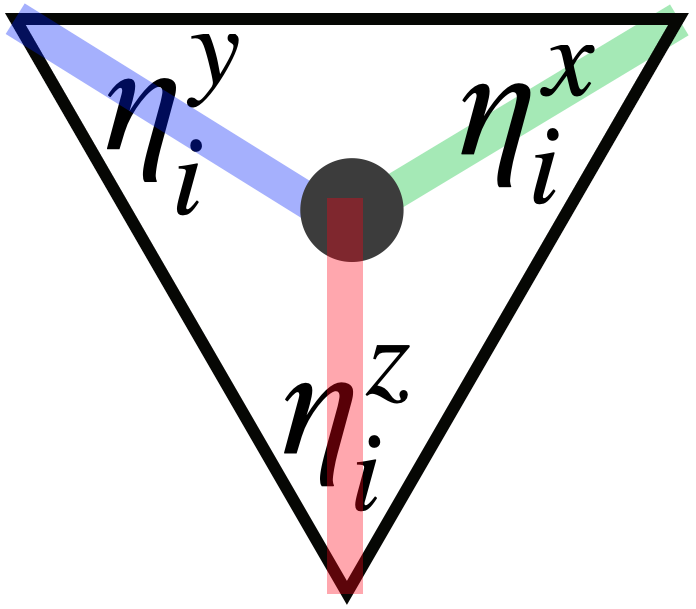} & i\in\text{B sublattice.}
\end{cases}\label{eq:etasublattice}
\end{equation}
We refer to this as the $\eta$-representation of the moments $\vec{S}_{i}$,
which allows us to easily extract the role of the spin component signs
$\eta_{i}^{\alpha}$ in the energy minimization process. For example
the signs of the two energy contributions from $\Gamma^{z}$ along
the $\langle12\rangle_{z}$ bond in Fig. \ref{zbondbasis} are $\text{sgn}\left(\Gamma^{z}S_{1}^{x}S_{2}^{y}\right)=-\,\eta_{1}^{x}\eta_{2}^{y}$
and $\text{sgn}\left(\Gamma^{z}S_{1}^{y}S_{2}^{x}\right)=-\,\eta_{1}^{y}\eta_{2}^{x}$,
which are minimized for general $a_{i},b_{i}$ when $\eta_{1}^{x}\eta_{2}^{y}=\eta_{1}^{y}\eta_{2}^{x}=1.$
In the pure $\Gamma$ limit all $\eta$-constraints may be minimized
by fixing the signs of an arbitrary site $i$ as $\left(\eta_{i}^{x},\,\eta_{i}^{y},\,\eta_{i}^{z}\right)\equiv\left(\eta_{1},\,\eta_{2},\,\eta_{3}\right)$
and distributing the signs by satisfying the $\eta$-constraints bond
by bond, introducing new $\eta$'s as necessary to parametrize the
signs of any leftover spin components. One may see that each plaquette
$p$ may be assigned an Ising variable $\eta_{p}$ by satisfying the
$\eta$-constraints along the plaquette's boundary. Furthermore, the
energy is insensitive to the value of $\eta_{p}$ as each contribution
squares to unity, so that the pure $\Gamma$ limit is equivalent to
an Ising gas on the triangular superlattice and exhibits an extensive
ground-state degeneracy \citep{Rousochatzakis_2017}. The combination of time-reversal and one of
the three global $\mathcal{R}_{\bar{\sigma}}$ transformations on
the $\Gamma$SL corresponds to flipping the signs of all $\eta_{p}$
which live on one of the plaquette sublattices $\bar{\sigma}\in\left\{ \bar{A},\bar{B},\bar{C}\right\} $
\citep{Rousochatzakis_2017,Samarakoon_2018}. For example, for the plaquettes
shown in Fig. \ref{gsleta}, applying $\mathcal{T}\cdot\mathcal{R}_{\bar{A}}$
on each moment gives
\begin{align}
\vec{S}_{1} & \rightarrow \left(-\eta_{a}|S_{1}^{x}|,\,S_{1}^{y},\,S_{1}^{z}\right), & \vec{S}_{4} & \rightarrow\left(S_{4}^{x},+\eta_{a}|S_{4}^{y}|,\,S_{4}^{z}\right),\nonumber \\
\vec{S}_{2} & \rightarrow\left(\,S_{2}^{x},\,S_{2}^{y},+\eta_{a}|S_{2}^{z}|\right), & \vec{S}_{5} & \rightarrow\left(\,S_{5}^{x},\,S_{5}^{y},-\eta_{a}|S_{5}^{z}|\right),\nonumber \\
\vec{S}_{3} & \rightarrow\left(S_{3}^{x},-\eta_{a}|S_{3}^{y}|,\,S_{3}^{z}\right), & \vec{S}_{6} & \rightarrow\left(+\eta_{a}|S_{6}^{x}|,\,S_{6}^{y},\,S_{6}^{z}\right),\label{eq:TRsigma}
\end{align}
and similarily for the $i'$ and $i''$ moments which will flip the
signs of $\eta_{b}$ and $\eta_{c}$, respectively. 

We add a finite Kitaev term and investigate the stability of the $\Gamma$SL.
For the $\langle61'\rangle_{x},\,\langle4'3''\rangle_{y},\text{ and }\langle2''5\rangle_{z}$
bonds shown in Fig. \ref{gsleta}, the Kitaev contributions to the
energy come with sign
\begin{align}
 & \text{sgn }E_{\langle61'\rangle_{x}}^{K}\hspace{0.14cm}=-\text{sgn }K\,\eta_{6}^{x}\,\eta_{1'}^{x}=\eta_{a}\,\eta_{b},\nonumber \\
 & \text{sgn }E_{\langle4'3''\rangle_{y}}^{K}=-\text{sgn }K\,\eta_{4'}^{y}\,\eta_{3''}^{y}=\eta_{b}\,\eta_{c},\nonumber \\
 & \text{sgn }E_{\langle2''5\rangle_{z}}^{K}\hspace{0.12cm}=-\text{sgn }K\,\eta_{2''}^{z}\,\eta_{5}^{z}=\eta_{c}\,\eta_{a},\label{eq:18kitaev}
\end{align}
and thus are minimized when $\eta_{a}\,\eta_{b}=\eta_{b}\,\eta_{c}=\eta_{c}\,\eta_{a}=-1$.
The three $\eta_{a,b,c}$ cannot be fixed simultaneously without violating
one of the $\eta$-constraints, which shows that perturbing the $\Gamma$SL
with Kitaev interactions is identical to the triangular Ising antiferromagnet
with interactions between next-nearest neighbor $\eta$-variables
\citep{wannier50,tanaka75,brandt86}. It also presents a clear demonstration
of the competition present between Kitaev and $\Gamma$ interactions
of opposite signs \citep{Liu_2021}. 

\begin{figure}
\includegraphics[scale=0.25]{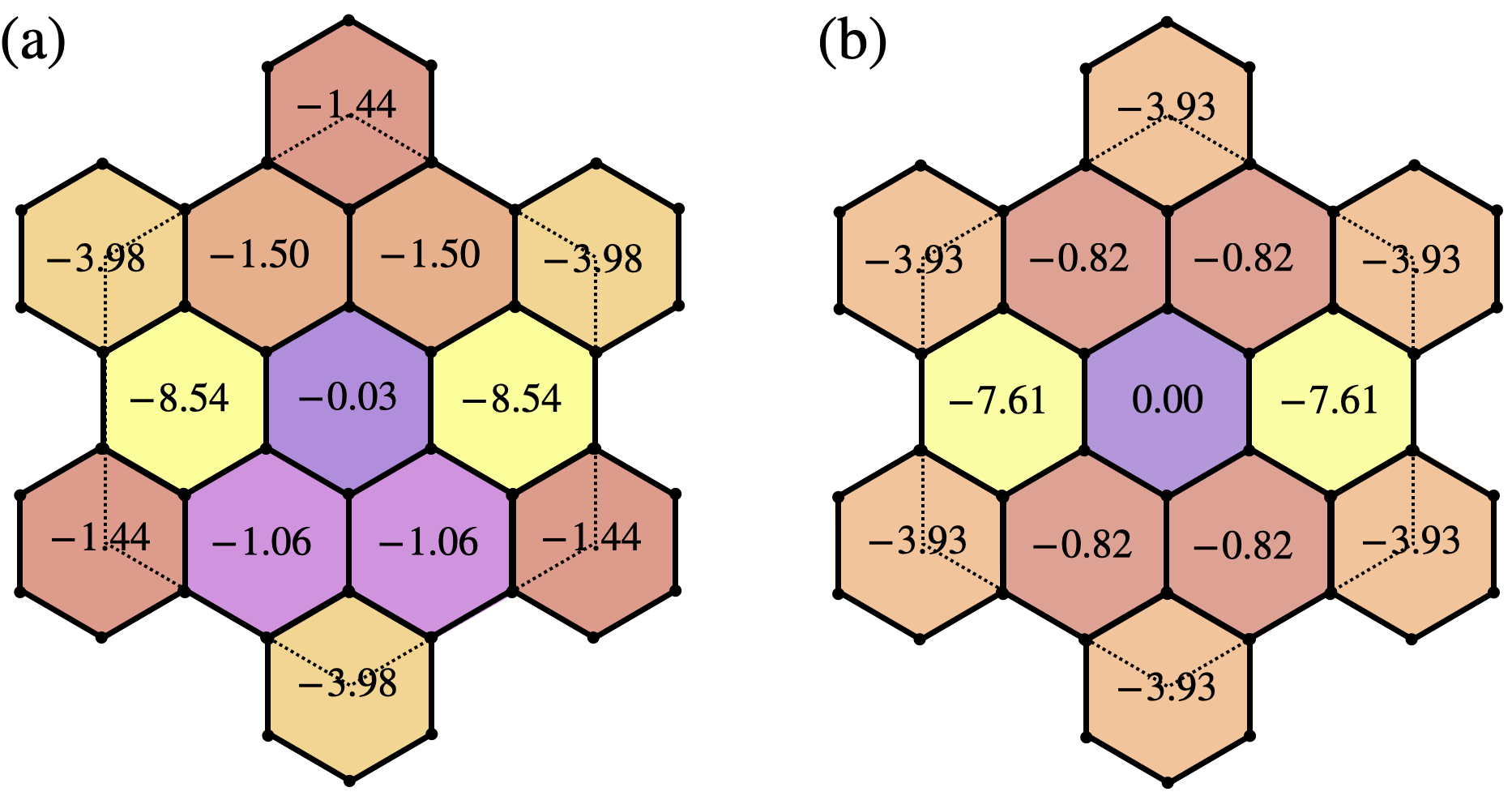}

\caption{The (a) $18^{\eta}$ order at $\left(\psi,\,g\right)=(0.4\pi,\,0.5)$
and (b) $18^{'}$ order at $\left(\psi,\,g\right)=(0.4\pi,\,0.75)$, with the values
of the plaquette flux $W_{p}/S^{6}$ shown in the center of each hexagon. The nine-plaquette unit cell is indicated by  lines. Note that
in (b) the $W_{p}$ are inversion symmetric about the zero-flux plaquette
as the $18'$ phase respects inversion symmetry.}

\label{wp-1}
\end{figure}

The ground state is obtained
when two-thirds of all $\eta_{p}$ carry one sign while the remaining
third carry the opposite sign \citep{wannier50}. 
For next-nearest neighbor interactions only, there are several configurations
of $\eta_{p}$ that minimize the energy \citep{brandt86}. However including the spin
magnitude $|S_{i}^{\alpha}|$ lifts this degeneracy and selects the
state with nine $\eta_{p}$, or $18$ sites, in the magnetic unit
cell. This is precisely the $18^{\eta}$ phase,
which is a subset of the classical degenerate ground states of the
$\Gamma$SL that is selected by FM $K$. Similar to the $\Gamma$SL
the $18^{\eta}$ phase exhibits well-defined plaquette fluxes $W_{p}\equiv2^{6}\prod_{i\in\partial p}S_{i}^{\text{out}(i)}\neq0$,
see Fig. \ref{wp-1}(a). 

The $18'$ phase is separated from $18^{\eta}$ phase via a second
order transition. It is similar to the $18^{\eta}$ phase with
eightfold-degeneracy but the spin patterns respect inversion symmetry.
We label the orientations of $18'_{i}$ as $i=x,y,z_{1},z_{2}$ and
we note that inversion maps each of $18'_{x,y}$ to itself, whereas
$18'_{z_{1}}$ maps to $18'_{z_{2}}$ and vice versa. The $18'$ phase
contains an ``idle'' plaquette with vanishing flux as shown in Fig.
\ref{wp-1}(b), and the six surrounding moments are pinned near the 
$[100]$, $[010]$, and $[110]$ axes. 
Note that these moments lie in the $xy$-plane, which reflects the increasing influence of
the $\Gamma$ dimer's $O(2)$ degeneracy discussed earlier as bond
anisotropy is increased. Thus the $18^{\eta}$ and $18'$ phases result
from the competing physics of the dimer and isotropic $\Gamma$ limits
in the presence of Kitaev interactions. 

\section{Effects of Quantum Fluctuations \label{sec:LSW}}

In this section we discuss the effects of quantum fluctuations on
the classical ground states by measuring the zero-point motion about
the ordered states. Using LSWT
the magnon gap is defined as $\Delta_{0}=\text{min }\omega_{\mathbf{k}}^{s}>0$
where $\omega_{\mathbf{k}}^{s}$ are the magnon dispersions and $s$
labels the sites of the magnetic unit cell. Increased quantum fluctuations
lead to a reduction of the moment magnitude
\begin{equation}
\braket{M}=S-\frac{1}{N}\sum_{i}\braket{a_{i}^{\dagger}a_{i}},\label{eq:reducedmoment}
\end{equation}
where $\braket{a_{i}^{\dagger}a_{i}}$ is the number of magnons per
site in the ground state $\ket{0}$ at $T=0$. 

\begin{figure}
\includegraphics[scale=0.28]{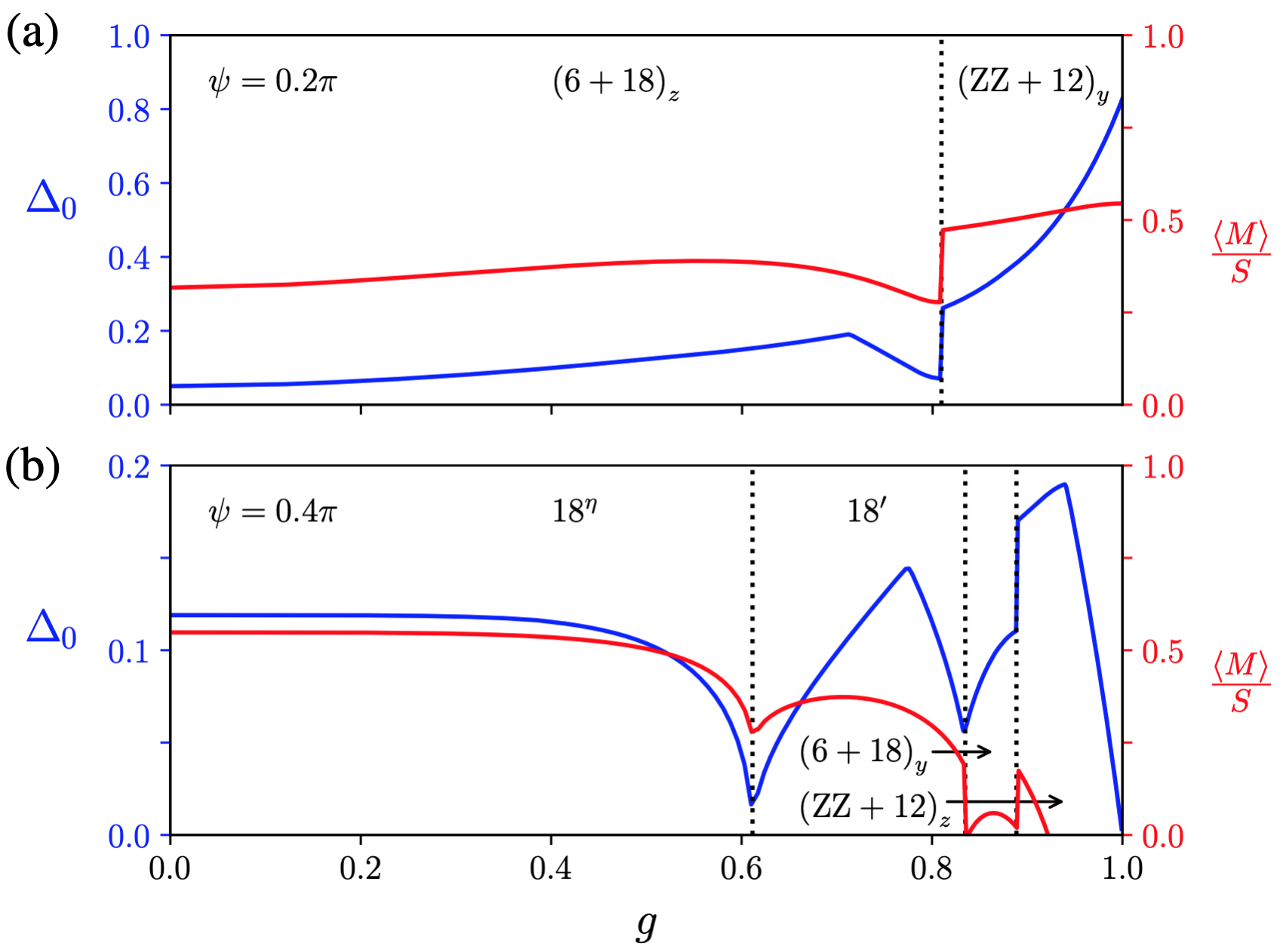}\caption{Reduced moment $\braket{M}/S$ (red) and magnon gap $\Delta_{0}$ (blue) as a function of $g$ for (a)
$\psi=0.2\pi$ and (b) $\psi=0.4\pi$. The classical phases stabilized
in each region are labeled and the phase boundaries are indicated
by the dashed black lines.}
\label{lswt}
\end{figure}
In Fig. \ref{lswt} we show two cuts where the bond anisotropy is
varied at fixed $\Gamma/K$ in the Kitaev dominant limit $\psi=0.2\pi$
$(\Gamma/|K|\sim0.73)$ and the $\Gamma$-dominant limit $\psi=0.4\pi$
$(\Gamma/|K|\sim3.08)$. We indicate the magnon gap $\Delta_{0}$
and the reduced moment $\langle M\rangle/S$ in blue and red, respectively.
$\langle M\rangle/S\sim0$ indicates that the classical order is unstable
due to quantum fluctuations. Throughout both cuts the moment is reduced
by more than 50\% indicating strong quantum effects in all phases.
However, the effects of anisotropy are qualitatively different between
the $K$ and $\Gamma$ regimes. In the Kitaev dominant region increased
anisotropy leads to a decrease in moment reduction, and the gap increases
due to the stability of the Ising easy-axis at the pure dimer limit.
On the other hand, in the $\Gamma$-dominant limit the gap goes to
zero as anisotropy is increased. This is due to the proximity of the
$O(2)$ symmetric $\Gamma$ dimer, which exhibits gapless excitations
within the $xy$-plane as discussed in Sec. \ref{sec:Hamiltonian-and-Phase}.
Interestingly, while the magnons are gapped away from this limit,
the reduced moment indicates that quantum fluctuations become strong
enough to completely destroy the magnetic order within LSWT.

The LSWT is valid up to $\mathcal{O}(1/S)$ and can return an unphysical
result when the number of magnons is large relative to $S$. This
occurs in the presence of low-lying flat magnon bands where the magnon-magnon
interactions cannot be ignored. In such a case the LSWT breaks down
and accurate calculation of observables requires a proper accounting
of effects beyond the single-magnon picture, which already tend to
be significant in non-collinear magnetic orders \citep{Consoli_2020,Smit_2020,Maksimov_2020}.
Nevertheless the reduced moment provides a general estimate of the
regions of the classical phase space which are most susceptible to
a quantum-disordered state \citep{fazekas99}, and we find that such
a phase may be stabilized in the $\Gamma$-dominant limit with moderate
anisotropy.

\section{Discussion and Summary\label{sec:Discussion}}
As we presented above, the K$\Gamma$ Hamiltonian with the bond anisotropy
$g$ is invariant under the $\mathcal{R}_{\bar{\sigma}}$ operators
of Eq. \eqref{eq:Rtransform}. However, when other interactions are
present the symmetry is generally broken. This includes the first
$(J)$ and third $(J_{3})$ nearest-neighbor Heisenberg interactions
and bond-dependent $\Gamma'$ interactions, which are relevant for
the description of Kitaev candidate materials \citep{Rau_2016,Takagi2019review,Motome_2020}.
For example, the $J$ term on the $\langle12\rangle_{z}$ bond in
Fig. \ref{zbondbasis} transforms under $\mathcal{R}_{\bar{A}}$ as
$J\vec{S}_{1}\cdot\vec{S}_{2}\rightarrow J\left(-S_{1}^{x}S_{2}^{x}-S_{1}^{y}S_{2}^{y}+S_{1}^{z}S_{2}^{z}\right)$,
and the $\Gamma'$ term as $\Gamma'\left[S_{1}^{x}S_{2}^{z}+S_{1}^{z}S_{2}^{x}+(x\rightarrow y)\right]$$\rightarrow$$\Gamma'\left[S_{1}^{x}S_{2}^{z}-S_{1}^{z}S_{2}^{x}-(x\rightarrow y)\right]$.
The external Zeeman field also breaks the symmetry. Thus the degeneracy
related to this symmetry in all the magnetic orders shown in this
study is lifted in the presence of any one or more of these terms.
This explains several results of previous studies including how a
FM $\Gamma'$ lifts the $(\text{ZZ}+12)_{z}$ degeneracy and selects
the $\text{ZZ}_{z}$ over the $\text{12}_{z}$ configurations \citep{rau2014trigonal,Gordon_2019,Lee2020Magnetic,luo2020spontaneous,hwang2021identification}.
Similarly the external magnetic field 
lifts the $\text{ZZ}+12$ and $6+18$ degeneracies near the Kitaev
region and selects the $\text{ZZ}$ and six-site orders, respectively
\citep{Chern_2020,Zhang_2021}.


{ \color{black}
We compare our results with previous
studies of the $K\Gamma$ model at the isotropic limit $g=0$. This was first explored in Ref. \citep{rau2014prl} where the classical limit shows the presence of an incommensurate spiral order along the $K\Gamma$ line using a single-$\mathbf{Q}$ variational ansatz. Further studies of the $K\Gamma$ line go beyond this approximation using classical Monte Carlo techniques \citep{lampenkelley2021fieldinduced,Chern_2020,sun2021effective,Liu_2021,rao2021machinelearned}. More recently it was found that the incommensurate spiral order is stabilized at low temperatures $T\sim0.1-0.2$ as exhibited by
the magnetic susceptibility and heat capacity \citep{sun2021effective}. In contrast, we obtain the LUC orders in this region of $g=0$
when $T\sim10^{-9}$ where thermal fluctuations are minuscule compared to the average interaction scale. We determine the energy of each phase to a high precision using the algorithm given in Ref. \citep{Chern_2020}, i.e., by annealing the cluster to this ultra-low temperature and then performing sweeps of the cluster where the moments are aligned with their local molecular fields, see Appendix \ref{sec:Simulation-Details} for details.

Numerical studies of the quantum model in the isotropic limit $g=0$ report} various quantum-disordered phases including a proximate KSL
(PKSL) \citep{Wang_2019-vmc,Wang_2020-vmc,Wang_2020_vmc}, a $\Gamma$SL
\citep{luo2020spontaneous,luo2021gapless}, and a nematic paramagnet
\citep{Lee2020Magnetic,Gohlke_2020}. Away from $g=0$, a disordered
region between the isotropic limit and the dimer phase, which rapidly
expands when $\Gamma>K$, was also reported \citep{Yamada_2020}. However,
it is not clear whether this is a true phase boundary or simply a
crossover region connecting the two phases. A series of multinode
gapless QSLs before entering a dimer phase at larger $g$ was found
using variational MC (VMC) \citep{Wang_2020-vmc}. A further VMC simulation
including the multi-$\mathbf{Q}$ orders around the $\Gamma$ region
with finite $g$ (i.e., $18^{\eta}$ and $18'$) would extend our current
knowledge on possible QSLs and their nature in this region.

Here, for the classical K$\Gamma$ model with $g=0$, we find four
different phases {\color{black}($\text{ZZ}+12$, $6+18$, $16+48$ and $18^{\eta}$)}  with
LUCs, and the LSWT shows that, in the $\Gamma$-dominant regime, the reduced moment $\langle M\rangle/S$
for fixed $\Gamma/K$ in Fig. \ref{lswt}(b) decreases as $g$
increases. This supports the disordered phases reported in Refs. \citep{Yamada_2020,Wang_2020-vmc}.
Furthermore if the magnetic ordering in the $18_{x,y,z}^{\eta}$ orientations
are destroyed by quantum fluctuations but the spontaneous $C_{3}$
symmetry breaking survives at $g=0$, it generates a nematic paramagnetic state.
We emphasize though this broken lattice-rotational symmetry
does not exclude a QSL.

Finally, an interesting proposal is a possible vison crystal spin
liquid near the $\Gamma$SL. In fact, one feature of the $\Gamma$SL
is significant correlations of the plaquette fluxes, which peak at
the $\Gamma$ and $K,K'$ points in the reciprocal space \citep{Saha_2019,luo2021gapless}.
A vison crystal spin liquid, which is magnetically disordered yet
exhibits a broken translational symmetry in the form of a long-range
$\braket{W_{p}W_{p'}}$ correlation function, may be stabilized in
the $\Gamma$-dominant region with moderate $z$-bond anisotropy,
which remains as a subject for future study. 

In summary, we have studied the classical $K\Gamma-g$ model to understand
the phases out of two competing frustrated interactions and the effects
of bond anisotropy on their competition. The pure Kitaev and $\Gamma$
models have classical spin liquids with macroscopic degeneracy, but
when they are both present, we found there exist several LUC phases
occurring via their competition. All the phases have the intrinsic
degeneracy related to a product of $\pi$ rotations around the plaquette,
where a subset of degenerate states can have a smaller magnetic unit
cell as a special case of a larger magnetic unit cell state. Near
the $\Gamma$-dominant region, we find 18-site magnetic unit cells
which retain their degeneracy in the presence of bond anisotropy unlike
the $C_{3}$ related degeneracy appearing in the Kitaev dominant region at $g=0$.
The bond anisotropy enhances quantum fluctuations in the $\Gamma$
dominant region which suggests that this region hosts a potential
QSL.

\begin{acknowledgments}
We would like to thank K. Chen, P. P. Stavropoulos, E. Z. Zhang, and J. Zhao for useful
discussions. We acknowledge support from the NSERC Discovery Grant
No. 06089-2016. H.Y.K also acknowledges support from CIFAR and the
Canada Research Chairs Program. Computations were performed on the
Niagara supercomputer at the SciNet HPC Consortium. SciNet is funded
by: the Canada Foundation for Innovation under the auspices of Compute
Canada; the Government of Ontario; Ontario Research Fund - Research
Excellence; and the University of Toronto.
\end{acknowledgments}

\appendix
\section{Monte Carlo Simulation Details\label{sec:Simulation-Details}}
\begin{figure}
\begin{centering}
\includegraphics[scale=0.2]{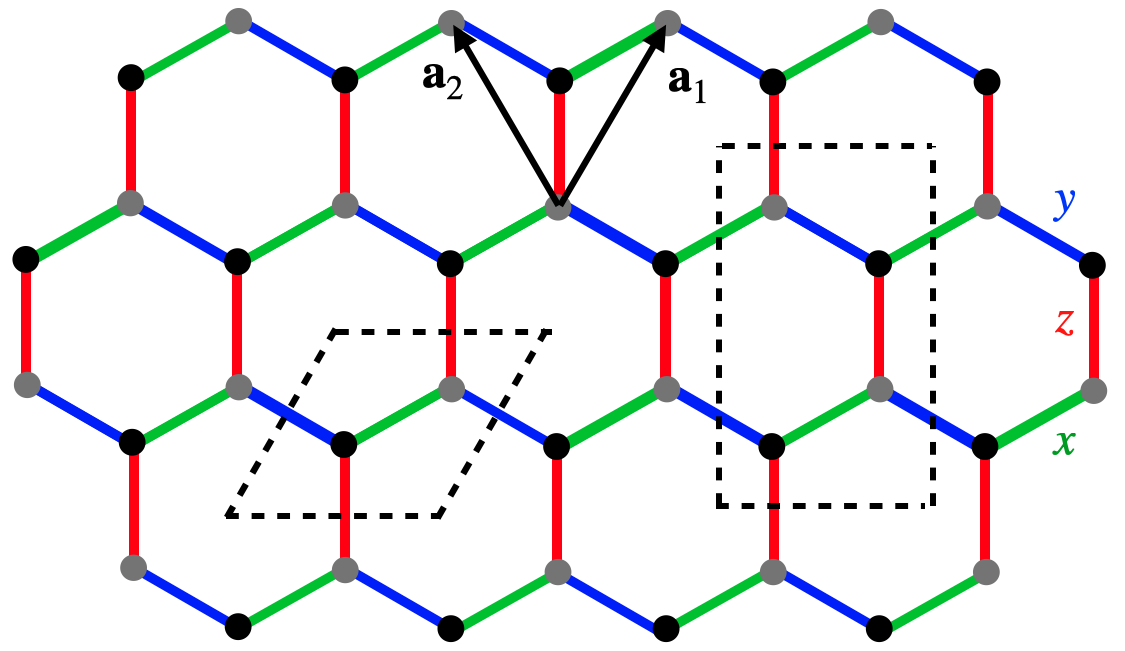}
\par\end{centering}
\caption{The two unit cells used to construct the Monte Carlo cluster, with
vectors $\mathbf{a}_{1}$ and $\mathbf{a}_{2}$ shown along with the
three bond types. From left to right: the rhombic unit cell with unit
cell vectors Eq. \eqref{eq:rhom2} and the rectangular unit cell with unit
cell vectors Eq. \eqref{eq:rect2}. The bond type $x,y,z$ are colored
green, blue, and red respectively.}

\label{unitcells}
\end{figure}
\begin{figure}
\begin{centering}
\includegraphics[scale=0.235]{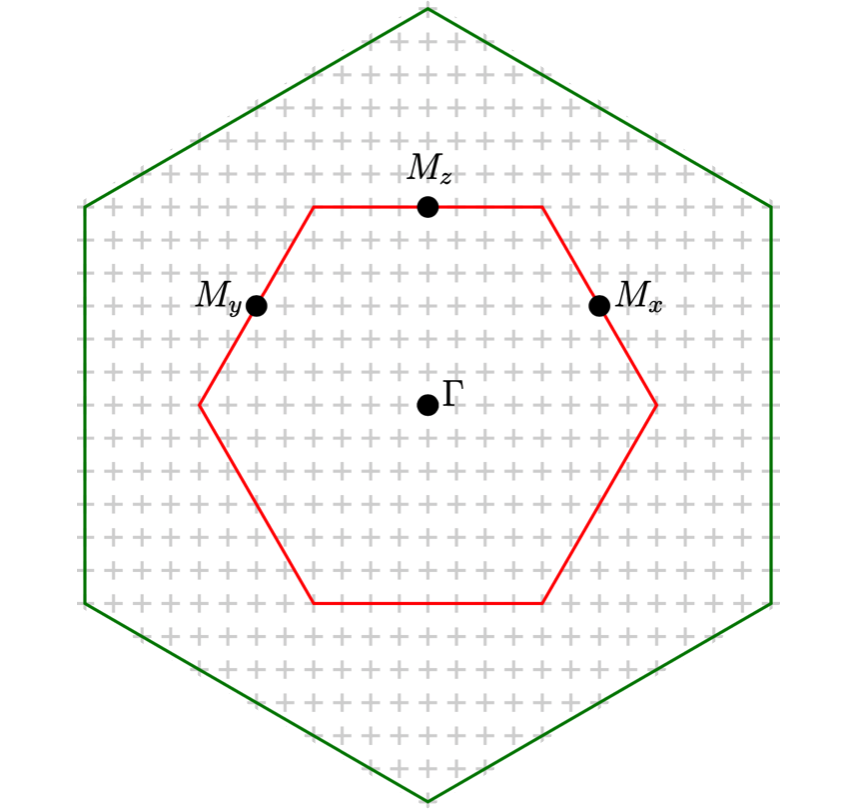}
\par\end{centering}
\caption{The accessible momentum points for the $N=4\times12\times6=288$ site
cluster constructed using the rectangular unit cell in Eq. \eqref{eq:rect2}.
The red and green hexagons correspond to the first and
second Brillouin zones, respectively. The three $M$ points
are shown and we use the label $x,y,z$ to distinguish different orientations
of the same magnetic order: see Appendix \ref{sec:SSFclassical}.}

\label{accessible}
\end{figure}
We perform the SAMC simulation on a finite size honeycomb cluster
with periodic boundary conditions. We parameterize the honeycomb sites
by placing them into unit cells located at $\mathbf{R}=\sum_{i}m_{i}\,\mathbf{T}_{i}$,
where $\mathbf{T}_{1},\,\mathbf{T}_{2}$ define the unit cell vectors
and $m_{1},m_{2}$ are integers. The honeycomb lattice contains $s\geq2$
sublattices within a unit cell, each forming its own sublattice: every
site $i$ can be labeled by three integers $i=(m_{1},\,m_{2},\,t)$
where $t=1,\ldots,s$. One choice for the geometry of the unit cell
is to use the two-site rhombic unit cell with translation vectors 
\begin{align}
\mathbf{T}_{1} & =\mathbf{a}_{1}-\mathbf{a}_{2}  =\left(1,\,0\right), \nonumber \\
\mathbf{T}_{2} & =\mathbf{a}_{1}  =\left(\frac{1}{2},\,\frac{\sqrt{3}}{2}\right),\label{eq:rhom2}
\end{align}
which is shown in Fig. \ref{unitcells} along with $\mathbf{a}_{1,2}$.
This choice produces a cluster that is commensurate with the $18$
site phases when $m_{1}$ and $m_{2}$ are multiples of 3. We may
also use the $4$-site rectangular unit cell with translation vectors
\begin{align}
\mathbf{T}_{1} & =\mathbf{a}_{1}-\mathbf{a}_{2}  =\left(1,\,0\right), \nonumber \\
\mathbf{T}_{2} & =\mathbf{a}_{1}+\mathbf{a}_{2}  =\left(0,\,\sqrt{3}\right),\label{eq:rect2}
\end{align}
which produces a cluster commensurate with the $(\text{ZZ}+12)_{z}$
order whenever $m_{1}$ is a multiple of $3$. Either unit cell may
be used to build the cluster as long as one ensures that the ordering
wavevectors of the classical states are accessible. For our purposes we use the $N=288$ site cluster with a rectangular unit cell, which accesses the $M,\,K/2,\text{ and }2M/3$ reciprocal points as shown in Fig. 10. We have also extended the cluster size to $N=720$ to check whether other large unit cell orders are stabilized.

After constructing the cluster we obtain the classical ground state
using the SAMC algorithm given in Ref. \citep{Chern_2020},
where Monte Carlo trials are performed for a finite temperature $T$
which is slowly tuned to zero. For our simulations we anneal according
to the cooling schedule $T_{i+1}=0.9\,T_{i}$ until the final temperature
$T_{f}=(0.9)^{200}\sim10^{-9}$ is reached. At each temperature
step $T_{i}$ of the simulation, we perform $N\times10^{5}$ Metropolis
trials where we choose a random moment, ``flip'' it so that it points
in a random direction, and accept the new configuration with probability
$\text{min}\left(1,\,e^{-\Delta E/T}\right)$ where $\Delta E$ is
the energy difference between the two states. When the final temperature
is reached, we further refine the energy by choosing a random moment
and aligning it with its local molecular field, and then repeating
this $N\times10^{4}$ times.

In order to obtain the phase diagram Fig. \ref{pd} we
first perform the SAMC on large clusters to resolve the possible classical
ground states, and then refine the energy of each phase by either
running the SAMC on small clusters or parametrizing the moments with
angles $(\phi_{i},\,\theta_{i})$ and minimizing the total energy
within the unit cell with respect to each angle. {\color{black} To accurately obtain the $T=0$ classical phase diagram it is important to determine each order's energy to full precision as the energy difference of the competing phases is within $\Delta E\sim\mathcal{O}(10^{-3})$. Otherwise at higher temperatures one may stabilize a mixture of the competing phases. For example, the ZZ, 6-site, and 16-site phases, which have SSF peaks along the $\Gamma-M$ lines in reciprocal space, are close in energy for $\psi/\pi\sim0.07-0.09$ along $g=0$. A mixture of the three phases would appear as an incommensurate order with SSF peaks that vary along $\Gamma-M$, but such a state remains higher in energy than the true classical ground state at $T=0$.}

\section{Magnetic Order of the K$\Gamma$ Classical Ground States\label{sec:SSFclassical}}
\begin{figure}[H]
\includegraphics[scale=0.31]{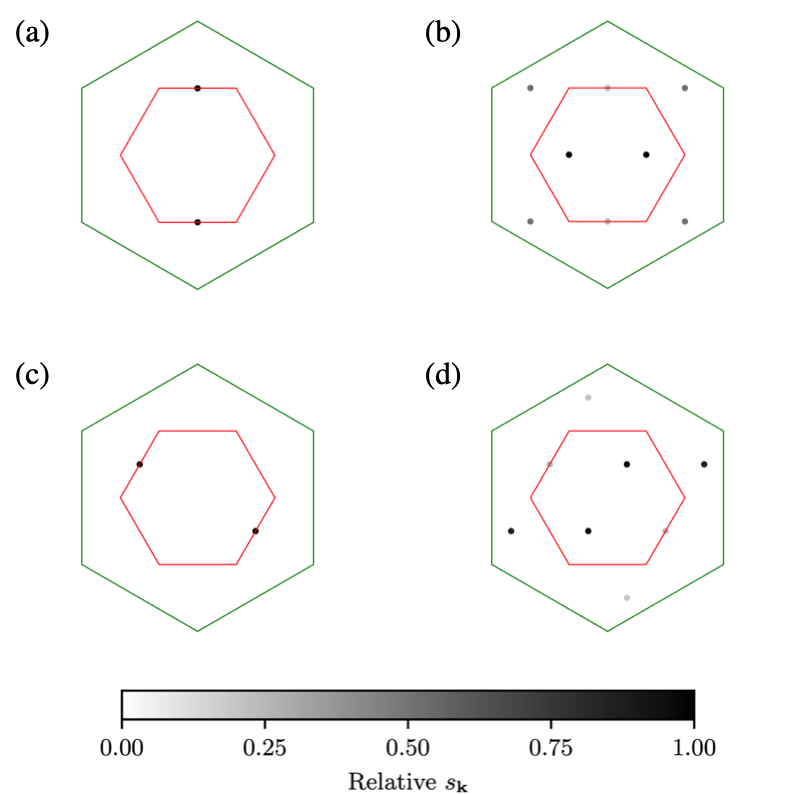}
\caption{
SSF of the (a) $\text{ZZ}_{z}$ and (b) $12_{z}$ patterns at $\left(\psi,g\right)=\left(0.4\pi,0.94\right)$, as well as the
(c) $\text{ZZ}_{y}$ and (d) $12_{y}$ patterns at $\left(0.15\pi,0.94\right)$ and $\left(0.1\pi,0.65\right)$, respectively.
The ZZ peaks are located at one of the three $M$ points, whereas the $12$-site peaks are located at one of three $\frac{1}{2}K,\frac{1}{2}K'$ pairs. The color of each ordering wavevector in Figs. \ref{ssfzz12}-\ref{ssf1648} denotes the relative intensity of each peak.
}
\label{ssfzz12}
\end{figure}

\begin{figure}[H]
\includegraphics[scale=0.31]{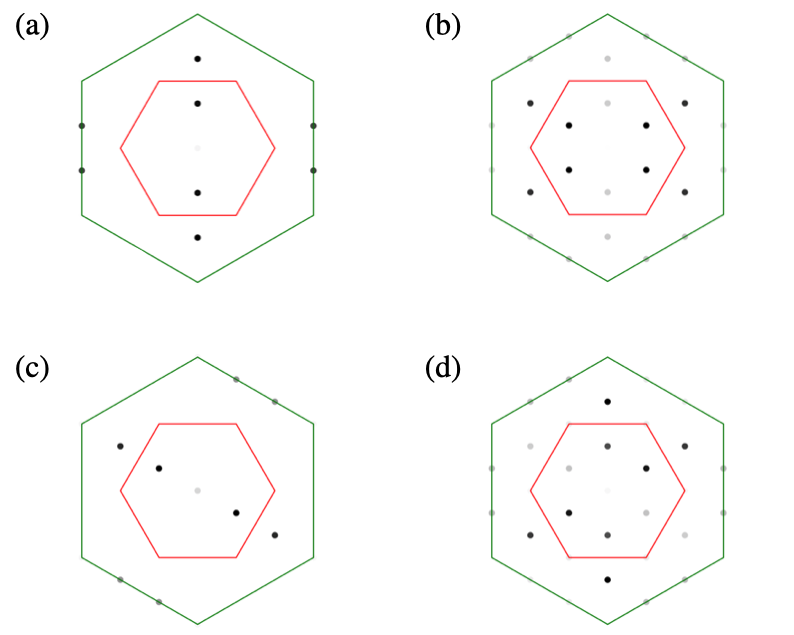}
\caption{
SSF of the (a) $6_{z}$ and (b) $18_{z}$ patterns at $\left(0.25\pi,0.84\right)$, as well as the
(c) $6_{y}$ and (d) $18_{y}$ patterns at $\left(0.35\pi,0.84\right)$.
The six-site peaks are located at one of three $\frac{2}{3}M,\frac{4}{3}M$ pairs, whereas the $18$-site peaks are located at multiple $\frac{2}{3}M,\frac{4}{3}M$ pairs.
}
\label{ssf618}
\end{figure}

\begin{figure}[H]
\includegraphics[scale=0.315]{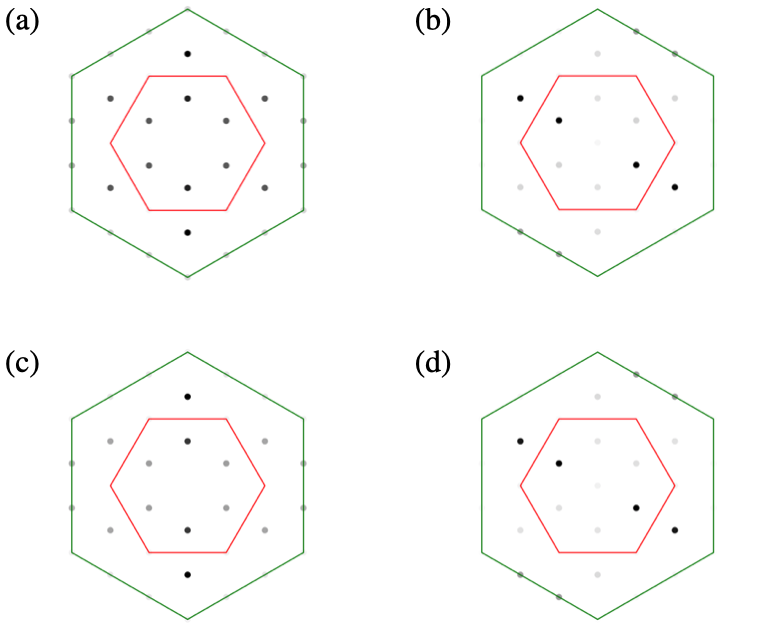}
\caption{
SSF of the (a) $18^\eta_V$ and (b) $18^\eta_y$ patterns at
$\left(0.4\pi,0.5\right)$, as well as the (c) $18'_{z_1}$ and (d) $18'_{y}$ phase at
$\left(0.4\pi,0.71\right)$. The ordering vectors are located at multiple $\frac{2}{3}M,\frac{4}{3}M$ pairs.
}
\label{ssf18etap}
\end{figure}

\begin{figure}[H]
\includegraphics[scale=0.31]{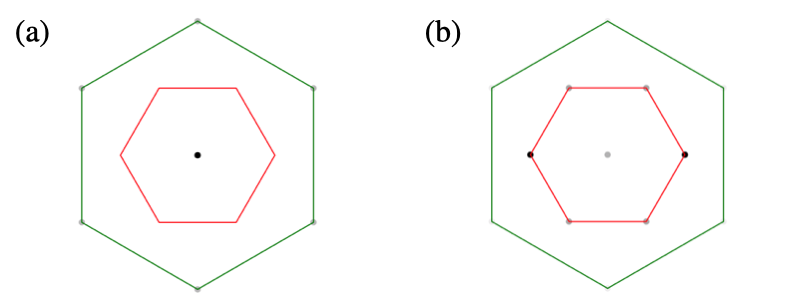}
\caption{SSF of the (a) $\text{FM}_{z}$ and (b) $120_{z}^\circ$ phase at
$\left(0.04\pi,0.5\right)$. The dominant SSF peak is at the $\Gamma$ point for the FM phase and one of three $K,K'$ pairs for the $120^\circ$ phase.}
\label{ssffm120}
\end{figure}

\begin{figure}[H]
\includegraphics[scale=0.31]{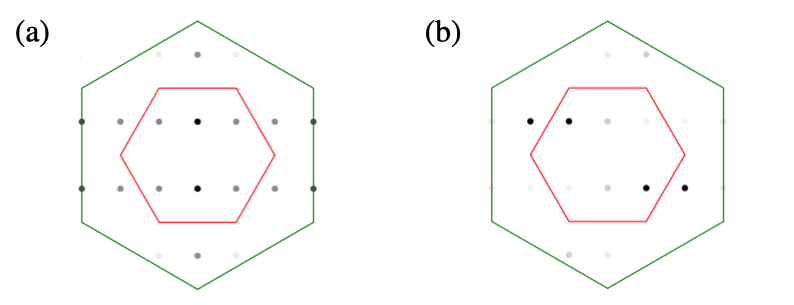}
\caption{SSF of the (a) $24_{z}$ and (b) $24_{y}$ patterns at
$\left(0.125\pi,0.57\right)$. The $24_z$ dominant SSF peaks are at $\frac{1}{2}M_z$ whereas for $24_y$ they are at $\left(\frac{1}{12},\frac{5}{12}\right), \left(-\frac{1}{12},\frac{7}{12}\right)$ in the $\left\{\mathbf{b}_{1},\mathbf{b}_{2}\right\}$ basis, where
 $\mathbf{b}_{1,2}$ satisfy $\mathbf{a}_i\cdot\mathbf{b}_j=2\pi\delta_{ij}$.}
 \label{ssf24}
\end{figure}

\begin{figure}[H]
\includegraphics[scale=0.17]{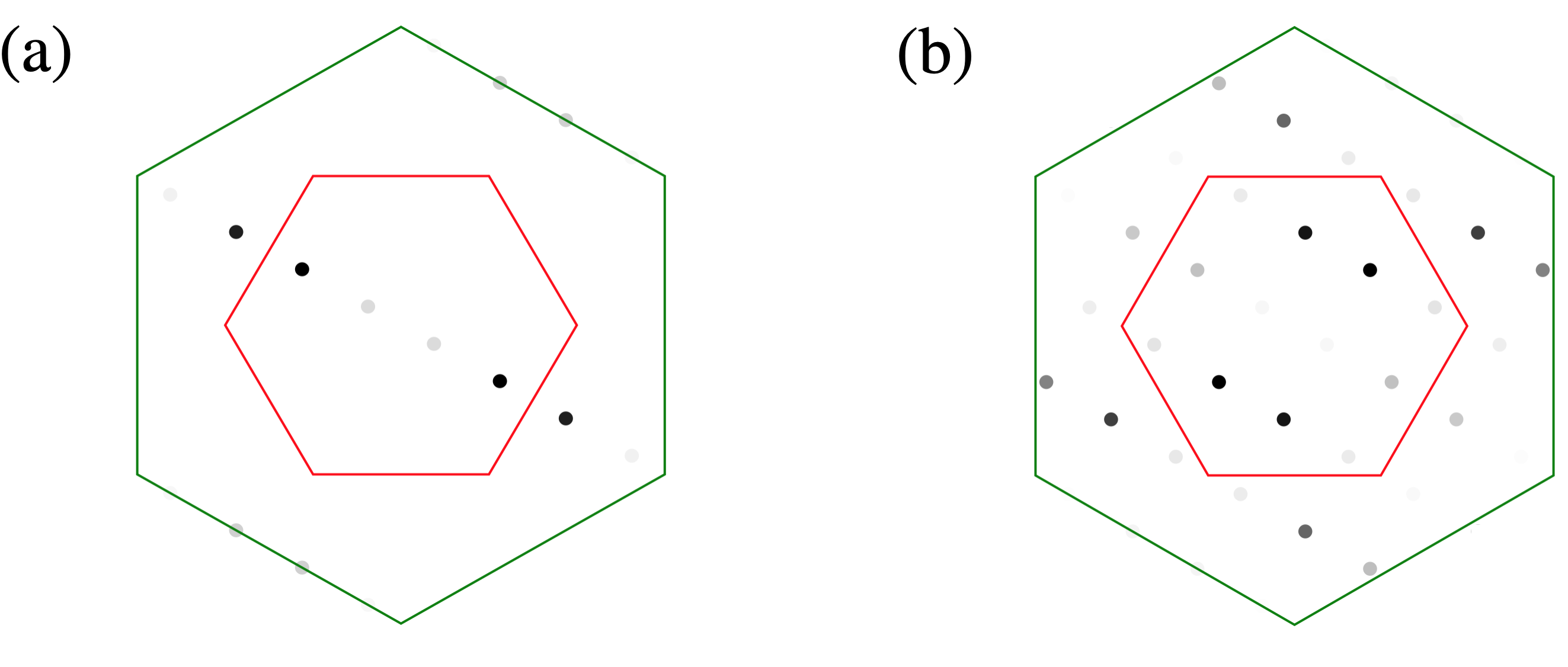}
\caption{ {\color{black}SSF of the (a) $16_{y}$ and (b) $48_{y}$ patterns at
$\left(0.14\pi,0.5\right)$. The $16_y$ dominant SSF peaks are at $\frac{3}{4}M_y, \frac{5}{4}M_y, $ whereas for $48_y$ they are at $\left(-\frac{1}{3},-\frac{1}{24}\right), \left(\frac{1}{3},\frac{7}{24}\right)$ in the $\left\{\mathbf{b}_{1},\mathbf{b}_{2}\right\}$ basis, where
 $\mathbf{b}_{1,2}$ satisfy $\mathbf{a}_i\cdot\mathbf{b}_j=2\pi\delta_{ij}$.}}
 \label{ssf1648}
\end{figure}

We present in Figs. \ref{ssfzz12}-\ref{ssf1648} the SSF patterns of the phases shown in the phase
diagram Fig. \ref{pd}. In each plot we show the first
and second Brillouin zones in red and green, respectively.
Different orderings, which share the same size of magnetic unit cell
but with different moment orientations, are distinguished by a subscript
$i=x,y,z$. These labels are assigned by the distribution of the SSF ordering
vectors about one of the three $\Gamma-M_{i}$ lines in Fig. \ref{accessible}.


\providecommand{\noopsort}[1]{}\providecommand{\singleletter}[1]{#1}%
%


\end{document}